\documentclass[conference]{IEEEtran}
\IEEEoverridecommandlockouts
\usepackage{algorithm}
\usepackage{array}
\usepackage[caption=false,font=normalsize,labelfont=sf,textfont=sf]{subfig}
\usepackage{url}
\usepackage{verbatim}
\usepackage{tikz}
\usepackage{circuitikz}
\usetikzlibrary{positioning}
\usepackage{dsfont}
\usepackage{comment}

\usepackage{cite}
\usepackage{amsmath,amssymb,amsfonts}
\usepackage{algorithmic}
\usepackage{graphicx}
\usepackage{textcomp}
\usepackage{xcolor}

\def\BibTeX{{\rm B\kern-.05em{\sc i\kern-.025em b}\kern-.08em
    T\kern-.1667em\lower.7ex\hbox{E}\kern-.125emX}}
    
\begin{document}

\title{Extended Sensitivity-Aware Reactive Power Dispatch Algorithm for Smart Inverters with Multiple Control Modes}

\author{\IEEEauthorblockN{Mohammad Almomani, Ahmed Alkhonain, and Venkataramana Ajjarapu}
\IEEEauthorblockA{\textit{Department of Electrical \& Computer Engineering, Iowa State University, Ames, IA, USA} \\
Emails: \{mmomani, ahmedkh, vajjarap\}@iastate.edu}
}

\maketitle

\begin{abstract}
The increasing integration of Distributed Energy Resources (DERs) in distribution networks presents new challenges for voltage regulation and reactive power support. This paper extends a sensitivity-aware reactive power dispatch algorithm tailored to manage smart inverters operating under different control modes, including PQ, PV, and Volt-Var (VV). The proposed approach dynamically optimizes reactive power dispatch and voltage setpoints, enabling effective coordination among distribution systems as a virtual power plant (VPP) to support the transmission network. The algorithm is applied to the IEEE 13-bus and IEEE-123 bus test systems, and its performance is validated by comparing results with OpenDSS simulations across various operating scenarios. \textcolor{black}{Results show that the maximum error in the voltages is less than 0.015 pu.}

\end{abstract}

\begin{IEEEkeywords}
Smart Inverters, Reactive Power Dispatch, Volt/VAR Control, Linear AC-OPF.
\end{IEEEkeywords}

\section{Introduction}
The rapid integration of Distributed Energy Resources (DERs) such as photovoltaics, wind turbines, and energy storage systems has significantly changed the operational landscape of modern power systems. Traditionally, large synchronous generators connected to the transmission system provided voltage support and reactive power management. However, with the increasing penetration of DERs in distribution networks, both Transmission System Operators (TSOs) and Distribution System Operators (DSOs) must collaborate to ensure voltage performance, particularly during contingencies or grid disturbances ~\cite{Sen1,FERC2222}.

The IEEE 1547-2018 standard provides guidelines for the interconnection and interoperability of DERs with electric power systems, emphasizing the need for coordinated reactive power management between TSOs and DSOs \cite{Sen1}. While this standard outlines the technical requirements for DER integration, it does not fully address the operational challenges that arise from the diverse types of DERs, including Grid-Forming (GFM) and Grid-Following (GFL) inverters. These inverter types have different characteristics and control strategies, complicating their integration into voltage support and reactive power dispatch frameworks \cite{Sen2,Der1}.

%Reactive power support from DERs has been identified as a crucial component in maintaining voltage performance. As more DERs are connected at the distribution level, the potential for these resources to provide voltage support in both normal and emergency conditions has increased. However, the traditional methods for reactive power dispatch, developed primarily for synchronous generators, are not directly applicable to DERs, especially in the presence of GFM and GFL inverters with distinct operational modes \cite{DER}.

One of the key challenges is that GFM inverters, which regulate voltage, and  GFL inverters, which control power injection based on grid voltage, require different approaches for reactive power dispatch. These smart inverters can operate as power-controlled sources (PQ buses), voltage-controlled sources (PV buses), Volt-VAR control mode (VV), Volt-WATT droop, WATT-PF, and WATT-VAR \cite{opendss, Sen1}.

Several recent studies have explored methods to optimize the reactive power dispatch of DERs in response to voltage regulation needs. A notable contribution is the \textbf{Sensitivity-Aware Reactive Power Dispatch} method, which prioritizes DERs based on their sensitivity to voltage at the substation \cite{Sen}. This approach minimizes the number of control signals required for dispatch and ensures that the most effective DERs are utilized first in providing voltage support. This framework is particularly beneficial for large networks where managing communication and control signals is a major challenge \cite{DER, Sen4}. While this method offers significant improvements in reducing communication complexity and reducing system losses, it primarily focuses on systems with PQ inverts' mode and does not fully address the unique requirements of other controllers.

In steady-state analysis involving different smart inverter controllers, traditional methods \cite{opf1, opf2} designate the most significant distributed generation (DG) unit as the slack bus. In contrast, distribution power flow analysis assigns other units as PV or PQ buses. However, this approach aligns differently with the decentralized nature of the distribution networks, where similarly sized units and droop-based control strategies prevent any single unit from stabilizing frequency and voltage. Moreover, it assumes the presence of a slack DG unit during islanded operation, which may not be feasible in practical scenarios.
Alternative approaches \cite{opf4, opf5} have been developed to address these limitations, offering power flow formulations specifically tailored for islanded case. These methods incorporate DG operating modes, including droop, PV, and PQ, represented through nonlinear equations for balanced and unbalanced systems. 
 Advanced models integrating additional controllers, such as virtual impedance and Virtual Synchronous Generators, are discussed in \cite{opf7, opf8}, which expand on primary and secondary control functions within nonlinear quasi-static power flow solutions. 

A significant gap in the literature is that most studies introduce smart invert modes to nonlinear power flow equations, limiting scalability in complex analyses. Our work advances this field by integrating smart inverter controllers within linear AC power flow equations considered in the dispatch algorithm. For instance, the approach in \cite{Sen} dispatches reactive power based on TSO requirements, without accounting for an optimized, autonomous strategy within distribution networks, thereby missing the potential for distribution networks to operate as virtual power plants (VPPs) with autonomous control. This paper addresses these limitations by proposing an extended dispatch algorithm that integrates diverse smart inverter controllers and dispatches Volt/VAR curves directly to individual DERs rather than solely meeting TSO reactive power requests, paving the way for optimized, decentralized control.

The contributions of this paper are twofold.
\begin{itemize}
    \item \textbf{Integrated smart inverter controllers into Linear OPF}  This work advances the current literature by incorporating smart inverter controllers into the Line-Dist Optimal Power Flow (OPF) formulation, enhancing scalability and efficiency in the dispatch algorithm.
    \item \textbf{Volt-Var curve dispatch as a VPP} Current reactive power dispatch methods, such as the approach in \cite{Sen}, are primarily based on meeting TSO requirements. However, there is a need for more optimized, autonomous dispatch strategies within distribution networks, which prevents these networks from fully realizing their potential as virtual power plants (VPPs) with decentralized control. This paper addresses this limitation by proposing an extended dispatch algorithm that integrates diverse smart inverter controllers and dispatches Volt/VAR curves directly to individual DERs, enabling optimized, decentralized control and enhancing the autonomy of distribution networks as VPPs.

\end{itemize}

The remainder of this paper is organized as follows: Section II \textcolor{black}{introduces the modeling of the smart inverter modes in the distribution networks}. Section III presents the proposed methodology, the control signals dispatch algorithm for reactive power and voltage set point. Section IV demonstrates the effectiveness of the proposed approach through numerical simulations, and Section V concludes the paper with a discussion of future work.

%\section{Background and Related Work}
\section{\textcolor{black}{Modeling smart inverter modes in distribution network}}
\subsection{\textcolor{black}{unbalanced Linear OPF}}

\textcolor{black}{The LinDistFlow \cite{LineDistFlow} model is a linearized approach to power flow that has been widely used to solve approximate optimal power flow (OPF) problems in balanced distribution networks efficiently. This model allows for rapid and accurate solutions to OPF problems, making it suitable for applications requiring scalable and computationally efficient methods. However, its original formulation is limited to balanced networks. To address this limitation, an extension called LinDist3Flow \cite{LineDist3Flow} has been developed for unbalanced distribution systems. LinDist3Flow adapts the linearized framework to handle three-phase, unbalanced networks, where voltage regulation and phase balancing are critical.}

\textcolor{black}{The power flow equation in LinDist3P is defined as:
\begin{equation}
\mathbb{Y} = R^{eq} \left( p^g - p^l  \right) + X^{eq} \left( q^g - q^l \right) + \mathbb{Y}_0 
\label{PF_Equation}
..\end{equation}}

\textcolor{black}{where: $R^{eq}$, $ X^{eq}$ are the constant three-phase impedance matrices as defined in \cite{LineDist3Flow}. Here, $\mathbb{Y}$ represents the nodal voltages (magnitude square), $p^g$ and $q^g$ are the real and reactive power generation, $p^l$ and $q^l$ are the real and reactive power loads. Note that equation (\ref{PF_Equation} )can account for voltage drops caused by line losses in the formulation of the impedance matrices \cite{Der27}. These matrices can be optimized from Non-Linear simulation also for more accuracy \cite{taheri2024optimized}.}

\subsection{\textcolor{black}{Voltage Support through VPP}}

\textcolor{black}{The TSO often treats the aggregated ADNs as a VPP to coordinate multiple distribution networks connected to the transmission network. This allows coordination between VPPs and other sources within the transmission system to ensure reliable market operation. In this setup, the TSO can directly request reactive power support from each VPP based on real-time transmission optimal power flow (T-OPF) to support the transmission network \cite{Sen}. Furthermore, this interaction between the TSO and the DSO can be designed to incorporate advanced control strategies, such as Volt-VAR control (VVC), for more intelligent and responsive coordination \cite{interactionTSO_DSO}.}

\textcolor{black}{A critical aspect of managing ADNs for voltage support is accurately determining their reactive power (VAR) capability. The approach in \cite{DER} introduces a method to aggregate the reactive power capability of distributed energy resources (DERs) at the substation level, providing the DSO with a clear view of the available reactive power that can be dispatched in response to TSO requests. }

\textcolor{black}{In this study, we utilize a simple droop-based Volt-VAR curves. At the VPP level, a droop-based Volt-VAR curve is applied at \(j\)-th VPP as \( Q_j^g = K_{q,j} (V_j - V_{\text{set},j}) \), where \( K_{q,j} \) is specified by the TSO considering the capability curves for each VPP. }

\textcolor{black}{The droop characteristic typically establishes a relationship between voltage magnitude (V) and reactive (Q) power. In LinDistFlow, the relationship is linear in terms of the square of the voltage magnitude (Y) and power (Q). Consequently, the droop characteristic appears nonlinear when represented in terms of \( Y \) in (\ref{PF_Equation}). To address this, we can decompose \( Y - Y_0 \) in (\ref{PF_Equation}) into \( (V - V_0)(V + V_0) \), where the dominant term is \( (V - V_0) \) when the voltage is close to 1. By approximating \( V + V_0 = 2 \), we can linearize the droop characteristic in terms of \( Y \) and power. Thus, the droop characteristic becomes:}

\begin{equation}
   Q_j^g = \frac{K_{q,j}}{2} (Y_j - Y_{\text{set},j})
 \label{vv_VPP}
\end{equation}

\textcolor{black}{ Here, \(Y_j, Y_{set,j}, Q_j^g\) denote the square of voltage magnitude, the set-point square voltage magnitude and the reactive power output of the VPP \(j\), respectively.}

\subsection{\textcolor{black}{Voltage Support through DERs}}

\textcolor{black}{Within the distribution network, smart inverters can operate in various modes, including \cite{opendss, Sen1}:}

\begin{itemize}
    \item \textcolor{black}{\textbf{PQ Mode:} Fixed active power (P) and fixed reactive power (Q), where the inverter outputs a constant P and Q.}
    
    \item \textcolor{black}{\textbf{PV Mode:} Fixed active power (P) and fixed voltage (V), maintaining a constant P while regulating voltage.}
    
    \item \textcolor{black}{\textbf{Volt-VAR Mode:} The inverter adjusts reactive power (VARs) based on one or two volt-VAR curves, depending on monitored voltages, current active power output, and the inverter's capabilities.}

\end{itemize}

\textcolor{black}{Similar to VPP, we utilize a simple droop-based Volt-VAR curves for the DERs. The reactive power \( q_i^g = k_{q,i} (v_i - v_{set,i}) \) represent the Volt-VAR curve for the \( i \)-th DER. Here, we assume that \( k_{q,i} \) is constants, and the DSO can control these DERs using the \( v_{\text{set},i} \) commands. We use capital letters to represent voltages and reactive power at the VPP level and lowercase letters for individual DERs in the ADN. Similarly, the Linearize relationship between reactive power and the voltage sequare at the DER is given by:}
\begin{equation}
    q_i^g = \frac{k_{q,i}}{2} (y_i - y_{set,i}), 
\label{vv_DER}
\end{equation}

\textcolor{black}{ Here, \(y_i, y_{set,i}, q_i^g\) represent the square of voltage magnitude, the target square voltage magnitude and the reactive power output in the DER \(i\).}

\section{Proposed Methodology}

\subsection{ Representation of DERs}
The representation of DERs in the proposed framework extends the traditional "LinDist3Flow algorithm to incorporate different controls of smart inverters. This modification enables the system to automatically adjust the reactive power (VAR) support and manage the voltage.

\textbf{Used symbols:}
\begin{itemize}
    \item $\mathbb{I}_n$: identity matrix with size $(n \times n)$
    \item $\mathbb{I}_1$: one column vector with size $(n \times 1)$
    \item $\mathbf{0}_n$, $\mathbf{0}_{n \times m}$: zero vector, matrix
\end{itemize}
We begin by considering a system with $N$ nodes. Let the PQ inverters be labeled from $1$ to $n_{pq}$, the PV inverters from $1$ to $n_{pv}$, the Volt-Var from $1$ to $n_{vv}$, and the non-generator buses from $1$ to $n_{nan}$. The following identification matrices are defined to categorize each type of source:
\[
\mathbf{ID}_{pq}, \mathbf{ID}_{pv}, \mathbf{ID}_{vv}, \mathbf{ID}_{nan}
\]
where the element at position $(i,j)$ in each matrix is $1$ if the $j$-th source is connected to node $i$, otherwise it is $0$. Based on these identification matrices, we define the incidence matrices (\(M_{pq}, M_{pv}, M_{vv}, M_{nan}\)) derived from the origin graph incidence matrix (M) \cite{DER} as follows: \(M_{*} = M^{-1} \times ID_{*}\) where * refer to pq, pv, vv, and nan. Similarly,  we define the equivalent resistance and inductance matrices for each category of source as follow: \(R_{*,eq} = -M^{-T}Z_{D}^p \mathbf{L} M_{*}\) and \(X_{*,eq} = -M^{-T} Z_{D}^q \mathbf{L} M_{*}\). Where  \( Z_{D}^q ,Z_{D}^p  \) are the constant three-phase impedance matrices as defined in \cite{LineDist3Flow} and \( \mathbf{L}\) is the loss factor diagonal matrix as defined in \cite{Der27}. The purpose of the identification matrices is to rearrange the system in certain way without changing the definition of matrices \(M,  Z_{D}^q ,Z_{D}^p, \mathbf{L} \). The overall aggregated resistance and inductance matrices for the system are then represented as:
\[
R_{r}^{eq} = \begin{bmatrix}
R_{vv} & R_{pq,eq} & R_{pv,eq} & R_{nan,eq}
\end{bmatrix}
\]
\[
X_{r}^{eq} = \begin{bmatrix}
X_{vv,eq} & X_{pq,eq} & X_{pv,eq} & X_{nan,eq}
\end{bmatrix}
\]

It is easy to see that these matrices are invertible as long as $R_{eq}$ and $X_{eq}$ from the original "LinDist3Flow" are invertible. Using these resistance and inductance matrices, defined \(K_r\) using \(R_r^{eq}\) and \(X_r^{eq}\) using equation (20) in \cite{DER} to consider voltage dependent load. So, The power flow equation is defined as:
\begin{equation}
\mathbb{Y} = K^{-1} \left[ R_r^{eq} \left( \mathbf{p} \\
^g - \mathbf{p}^l a_0 \right) + X_r^{eq} \left( \mathbf{q}^g - \mathbf{q}^l a_0 \right) + t_r \mathbb{Y}_0 \right]
\label{PF_Equation2}
\end{equation}

where $a_0$ is the load coefficient to represent the partition of constant power load from the total load. 

Substitute (\ref{vv_DER}) in (\ref{PF_Equation2})  after rearrangement it can be written in compact form as follow: 
\begin{equation}
    \mathbb{Y}= A_{sys} \mathbb{X} + B_{sys}
    \label{eq:dispatch_matrix2}
\end{equation}
Where
\[ \mathbb{Y=}
\begin{bmatrix}
\mathbb{Y}_{vv}&
\mathbb{Y}_{pq} &
\mathbf{q_{pv}} &
\mathbb{Y}_{nan}
\end{bmatrix} ^T , 
\quad 
\mathbb{X}=\begin{bmatrix}
t_r & \mathbb{Y}_{vv,set} &
\textcolor{black}{\mathbf{q_{pq}}} &
\mathbb{Y}_{pv}
\end{bmatrix}^T
\]

\textcolor{black}{ Where \( B_{sys} \) and \( A_{sys} \) contain all constant terms. \(\mathbb{Y}_{vv}\), \(\mathbb{Y}_{pq}\), \(\mathbb{Y}_{pv}\), and \(\mathbb{Y}_{nan}\) are vectors representing the squared voltage magnitudes at buses with DERs operating under VV mode, PQ mode, PV mode, and at buses without DERs, respectively. \( q_{pv} \) and \( q_{pq} \) denote the reactive power outputs of DERs in PV mode and PQ mode, respectively. \(\mathbb{Y}_{vv,set}\) represents the squared voltage set points for DERs operating in VV mode.}

\subsection{Control Signals and Reactive Power Dispatch}

To simplify the next steps, we define the rows and blocks partition of \( A_{\text{sys}} \) and \( b_{\text{sys}} \) referring to $vv$, $pq$, $pv$ and non-generator buses as follows:
{ \small \[
A_{\text{sys}} = \begin{bmatrix}
A_{r1} \\
A_{r2} \\
A_{r3}\\
A_{r4}
\end{bmatrix}
=
\begin{bmatrix}
A_{11} & A_{12} & A_{13} & A_{14} \\
A_{21}& A_{22} & A_{23} & A_{23}\\
A_{31}& A_{32} &A_{33} & A_{34}\\
A_{41} & A_{42} & A_{43} &  A_{44} 
\end{bmatrix}, \quad 
b_{\text{sys}} = \begin{bmatrix}
b_{1} \\
b_{2} \\
b_{3} \\
b_{4}
\end{bmatrix}
\]}
\begin{itemize}
    \item  Reactive power flow can be derived using equation  incidence matrices as follow \cite{DER} : 
\[
\resizebox{0.47\textwidth}{!}{$
Q_{flow} = M_{vv} q_{vv} + M_{pq} q_{pq} + M_{pv} q_{pv} - M^{-1} q_l
$}\]
Rearranging this equation: 

\begin{equation}
   Q_{flow} =A_{Q_{flow}} \mathbb{X}+B_{Q_{flow}}
   \label{eq:q_flow2}
\end{equation}

Where 
\[A_{Q flow}= [\frac{k_q}{2} M_{vv}, M_{pv}, M_{pq}]
\begin{bmatrix}
A_{11} & A_{12}-I & A_{13} & A_{14} \\
A_{31}& A_{32} &A_{33} & A_{34}\\
0& 0 & I &  0 
\end{bmatrix}
\]

\[
B_{Q_{flow}} = \frac{k_d}{2}M_{vv} b_{1} + M_{pq}b_{3}- M^{-1} Q_l
\]
Based on this arrangement, the first three rows of A and B represent the reactive power flow at the substation (\(Q_t\)) which are divided into four parts as follow: 
\[
A_{Q_{flow}}= 
\begin{bmatrix}
    A_{Q_{flow1}} & A_{Q_{flow2}} & A_{Q_{flow3}} & A_{Q_{flow4}}
\end{bmatrix}
\]
which represent the vectors that corresponding to each part of \(\mathbb{X}\). 

\item Voltages can be rewritten from equation \ref{eq:dispatch_matrix2} as follow: 

\begin{equation}
\begin{bmatrix}
Y_{vv} \\
Y_{pq} \\
Y_{nan} 
\end{bmatrix}
=
\begin{bmatrix}
A_{r1} \\
A_{r2} \\
A_{r4}
\end{bmatrix}
\mathbb{X}
+
\begin{bmatrix}
b_{1} \\
b_{2}  \\
b_{4}
\end{bmatrix}=A_v \mathbb{X}+B_v
\end{equation}

\end{itemize}

\subsubsection{Sensitivity-Aware Dispatch (objective Function) }
The dispatch takes into account different inverters, ensuring that the most sensitive nodes contribute the most to voltage support. The objective function is defined similar to the sensitivity in \cite{Sen}. Based on previous relations we can write the objective function in matrix form:
\begin{equation}
\min \left( W_1 Y_{vv,set} + W_2 q_{pq} + W_3 Y_{pv} \right)
\label{eq:obj}
\end{equation}

where:
{\small
\(
W_1 = \mathbb{I}_1 - \frac{\partial Q_t}{\partial q_{vv}}= \mathbb{I}_1- \frac{1}{d}A_{Q_{flow2}} \left[\frac{k_d}{2} (A_{12}-I)\right]^{-1}
\),
\(
W_2 = \mathbb{I}_1 - \frac{\partial Q_t}{\partial q_{pq}}=\mathbb{I}_1- \frac{1}{d} A_{Q_{flow3}}
\),
\(
W_3 = \mathbb{I}_1 - \frac{\partial Q_t}{\partial q_{pv}}=\mathbb{I}_1-\frac{1}{d} A_{Q_{flow4}}A_{34}^{-1} 
\) and
{\small \(
\mathbf{d} = A_{Q_{flow2}} \left[\frac{k_d}{2} (A_{12}-I)\right]^{-1} \times \mathbb{I}_1 +  A_{Q_{flow4}}A_{34}^{-1} \times \mathbb{I}_1 +A_{Q_{flow3}} \times \mathbb{I}_1 
\)}
}

\subsubsection{Constraints}

The constraints in matrix form are defined as follows:

\begin{itemize}
    \item \textbf{Reactive power balance at the substation:}
    \[
    \mathbb{I}_{\text{diff}} A_{Qflow} \mathbb{X} < \epsilon - \mathbb{I}_{\text{diff}} B_{Qflow}
    \]
    where \( \mathbb{I}_{\text{diff}} \) is \textcolor{black}{a constant matrix} used to find the difference between the phases (\( |Q_a - Q_b|, |Q_b - Q_c| \)) at the substation, and \( \epsilon \) is the allowable tolerance.

    \item \textbf{Upper/Lower Voltages:}
    \[
    Y_l - B_v < A_v \mathbb{X} < Y_h - B_v
    \]
    where \( Y_h \) and \( Y_l \) are the high and low voltage square limits (1.05\(^2\), 0.95\(^2\)).
    
    \item \textbf{Maximum Power in Distribution Lines:}
    \[
    A_{Qflow} \mathbb{X} < q_{\text{max}} - B_{Qflow}
    \]
    
    \item \textbf{Maximum Reactive Power :}
    for PV inverters:
    \[
    -b_{r3} - \sqrt{S^2 - p_g^2} \leq A_{r3} \mathbb{X} \leq -b_{r3} + \sqrt{S^2 - p_g^2}
    \]
    for VV inverters:
     \[
    | \frac{k_d}{2}(A_{r1}-[0 \quad I \quad 0 \quad 0]) \mathbb{X} +\frac{k_d}{2} b_{r1}| \leq \sqrt{S^2 - p_g^2}
    \]

    \item \textbf{VPP Volt-Var Curve:}
    \[
    A_{Qflow} \mathbb{X} + b_{Qflow} = \frac{K_{q}}{2} (Y-Y_{set})
    \]
\end{itemize}

% \end{comment}

\textcolor{black}{ The sensitivity-aware dispatch is real time disaggregation for VPP Volt-Var curve into individual DERs based on the previous objective function, where \( W_1 \), \( W_{2} \), and \( W_3 \) are constant vectors. All constraints are linear inequalities, except for the VPP Volt-VAR curve, which is a linear equality constraint. The dispatch algorithm proceeds in two main steps. First, the matrices \( A_{sys} \) and \( B_{sys} \) are calculated. \( A_{sys} \) primarily depends on the system topology, while \( B_{sys} \) is determined by the operating point, including load values and the maximum power generated by DERs. Once these constant matrices are established, the problem is reformulated into a standard linear optimization problem, ensuring efficient computation.}  

\section{Simulation Results}

\begin{figure}
    \centering
    \includegraphics[width=0.6\linewidth]{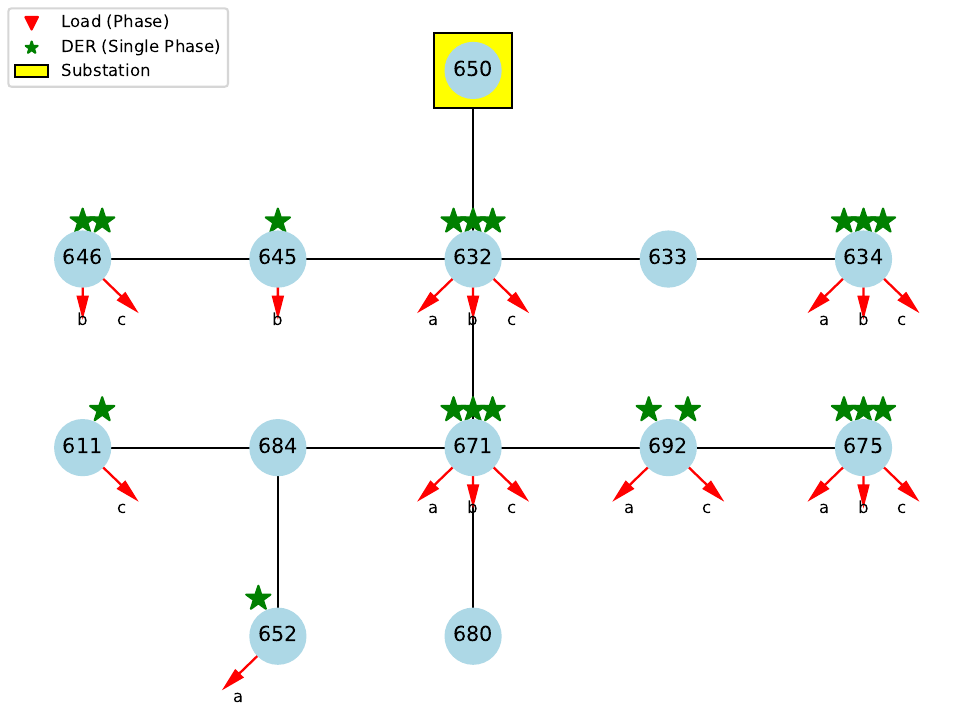}
    %\caption{IEEE-13 bus test system with DERs}
\caption{\textcolor{black}{IEEE 13-Bus Test System: Arrows Represent Per-Phase Loads, Stars Indicate Smart Inverters, and the Substation is Located at Node 650.}}
    \label{fig:IEEE13}
\end{figure}

The proposed extended sensitivity-aware dispatch algorithm was tested on the IEEE 13-bus test system \cite{IEEE13}, which includes various smart inverter controllers distributed across different nodes, Figure \ref{fig:IEEE13}. 
\textcolor{black}{As shown in Figure \ref{fig:IEEE13}, the system comprises 19 loads, with some connected to three-phase buses (buses 634, 671, and 675) and others connected to single-phase or two-phase buses. The transmission network is linked through bus 650, which represents the substation. The green stars indicate the locations of the DERs, which in this case are located with the loads. The position of each star corresponds to the phase to which the DER is connected: from left to right, the phases are \( a \), \( b \), and \( c \).}

In this study, we evaluated four scenarios: \textbf{PQ mode:} All DERs operate in PQ control mode, \textbf{PV mode:} All DERs operate in PV control mode,  \textbf{VV mode:} All DERs operate in VV control mode, and \textbf{Mixed mode:} DERs at buses 634 and 675 operate in VV control mode, the DER at bus 671 operates as a PV mode, and the remaining DERs operate in PQ control mode. The OpenDSS platform was used to validate the accuracy of the LineDist OPF models. First, system parameters (R, X) were optimized using OpenDSS in a manner similar to \cite{taheri2024optimized}. The distribution network was assumed to connect to the transmission system, along with other distribution networks, at the substation.

In this study, we configured a VPP linear Volt-Var curve, such that the distribution system does not provide any reactive power when the substation voltage is 1.0 p.u., but supports 2 MVAr of reactive power when the substation voltage is 0.95 p.u. At the DER level, we fixed the Volt-Var droop gain, ensuring that DERs provide maximum reactive power when the measured voltage is 0.1 p.u. below the setpoint voltage. The dispatch algorithm was then used to optimize the $V_\text{set}$ value for each Volt-Var DER.

\begin{figure}
    \centering
    \includegraphics[width=0.75\linewidth]{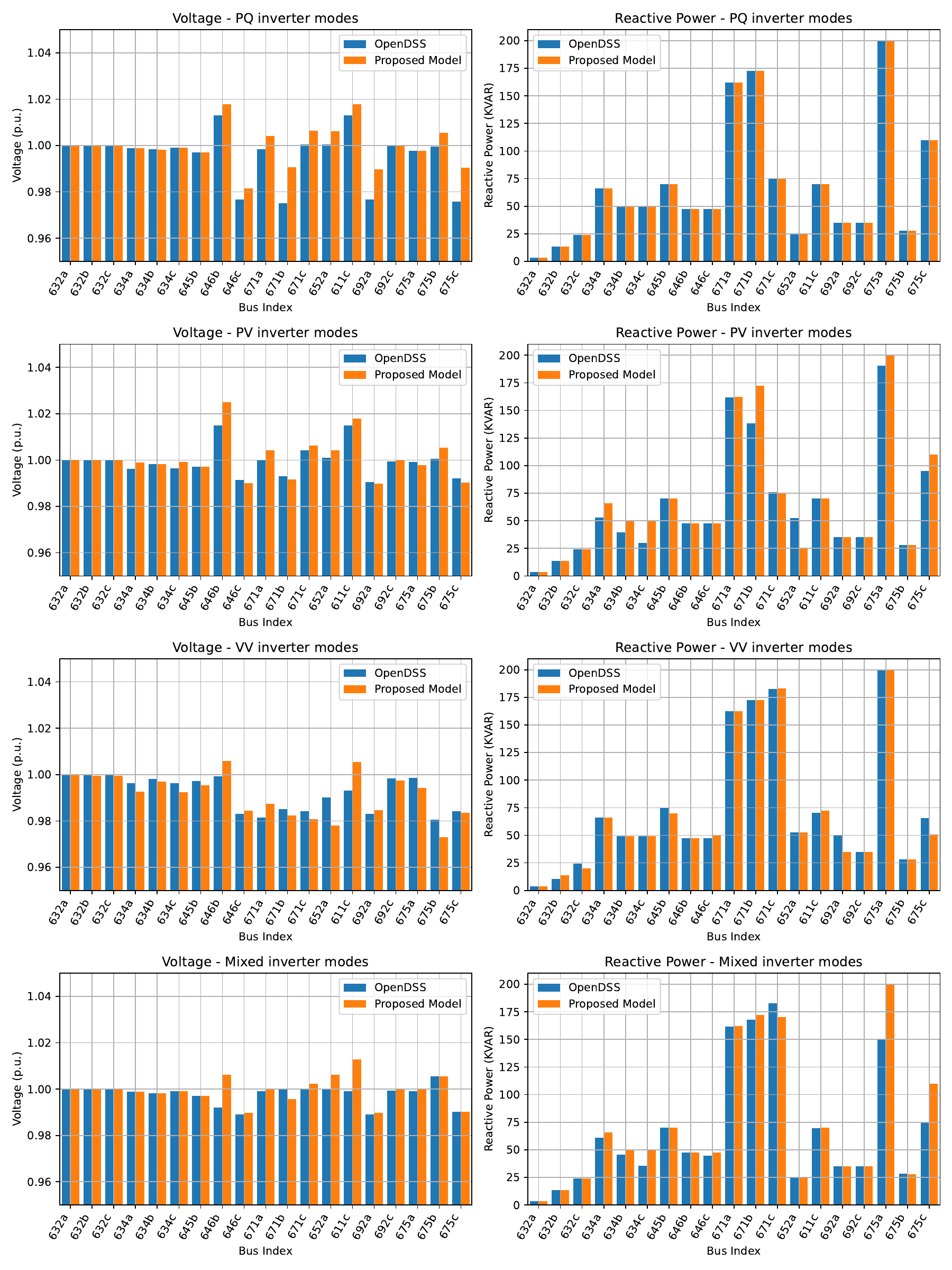}
    \caption{Comparison of Voltage Profiles and Reactive Power Across Different Inverter Modes for OpenDSS and Proposed Models. }
    \label{results}
\end{figure}

Figure \ref{results} illustrates the error between OpenDSS and the proposed method across the four scenarios. The first row of the figure shows the voltage and reactive power for the PQ mode. The reactive power closely matches between OpenDSS and the proposed linear system, while the voltage error is depicted in the upper left subplot. The maximum voltage deviation between OpenDSS and LineDist was 0.015 p.u. in the PQ mode, 0.01 p.u. in the PV mode (second row), 0.012 p.u. in the VV mode (Third row), and 0.014 p.u. in the mixed controller mode (Fourth row). The L1 norm of the voltage (pu) error and the L1 norm of the reactive power (KVAr) error (normalized by the total reactive power) are summarized in Table \ref{tab:L1NormComparisonUpdated}.

\begin{table}
\centering
\caption{Comparison of L1-Norm values for different cases}
\begin{tabular}{|c|c|c|c|c|}
\hline
\textbf{Case} & \multicolumn{2}{c|}{\textbf{IEEE 13-Bus}} & \multicolumn{2}{c|}{\textbf{IEEE 123-Bus}} \\ \hline
              & V (\%)   & Q (\%)   & V (\%)   & Q (\%)   \\ \hline
PQ            & 0.407           & 0                     & 0.52           & 0                     \\ \hline
PV            & 0.1            & 8.045                 & 0.17            & 7.67                 \\ \hline
VV            & 0.259          & 5.99                  & 0.273          & 5.34                  \\ \hline
Mixed         & 0.15           & 2.198                 & *           & *                 \\ \hline
\end{tabular}
\label{tab:L1NormComparisonUpdated}
\end{table}

\textcolor{black} {The algorithm, leveraging linear programming, ensures scalability to larger systems. When applied to the IEEE 123-bus system, it determined the optimal solution in under 0.5 seconds. The accuracy of the first three modes for the IEEE 123-bus system is presented in Table \ref{tab:L1NormComparisonUpdated}.}

\section{Conclusion and Future Work}

The extended sensitivity-aware reactive power dispatch method successfully manages different smart inverters functionalities in unbalanced distribution systems. Additionally, the use of linear programming ensures scalability, allowing the algorithm to solve the IEEE-123 bus test case in under 0.5 seconds. The algorithm's accuracy is validated by comparing its results with those from a nonlinear simulator, showing acceptable performance. \textcolor{black}{Here, the \(L_1\) norm of the error in the reactive power is around 8\% and less than 0.5\% for the voltage. Future work will involve incorporating more complex load models, such as open delta and constant current loads, and load/source uncertainty.}

\section*{Acknowledgment}
This work was supported by the DOE through CyDERMS project (DOE CESER DE-FOA-0002503 award DE-CR0000040)

\bibliographystyle{IEEEtran}
\bibliography{ref}

\begin{thebibliography}{10}
\providecommand{\url}[1]{#1}
\csname url@samestyle\endcsname
\providecommand{\newblock}{\relax}
\providecommand{\bibinfo}[2]{#2}
\providecommand{\BIBentrySTDinterwordspacing}{\spaceskip=0pt\relax}
\providecommand{\BIBentryALTinterwordstretchfactor}{4}
\providecommand{\BIBentryALTinterwordspacing}{\spaceskip=\fontdimen2\font plus
\BIBentryALTinterwordstretchfactor\fontdimen3\font minus \fontdimen4\font\relax}
\providecommand{\BIBforeignlanguage}[2]{{%
\expandafter\ifx\csname l@#1\endcsname\relax
\typeout{** WARNING: IEEEtran.bst: No hyphenation pattern has been}%
\typeout{** loaded for the language `#1'. Using the pattern for}%
\typeout{** the default language instead.}%
\else
\language=\csname l@#1\endcsname
\fi
#2}}
\providecommand{\BIBdecl}{\relax}
\BIBdecl

\bibitem{Sen1}
IEEE, ``Ieee standard for interconnection and interoperability of distributed energy resources with associated electric power systems interfaces,'' \emph{IEEE Std 1547-2018 (Revision of IEEE Std 1547-2003)}, pp. 1--138, 2018.

\bibitem{FERC2222}
R.~M. Tapio and A.~C. Orrell, ``Ferc order no. 2222 and considerations for distributed wind,'' Pacific Northwest National Laboratory (PNNL), Richland, WA (United States), Tech. Rep., 2023.

\bibitem{Sen2}
{NREL}, ``Renewable electricity future | scenario viewer,'' \url{https://scenarioviewer.nrel.gov/}, accessed: 2024-10-13.

\bibitem{Der1}
L.~Bao, Z.~Huang, and W.~Xu, ``Online voltage stability monitoring using var reserves,'' \emph{IEEE Transactions on Power Systems}, vol.~18, no.~4, pp. 1461--1469, 2003.

\bibitem{opendss}
\BIBentryALTinterwordspacing
{Electric Power Research Institute (EPRI)}, ``Opendss - open distribution system simulator,'' 2023, accessed: 2024-11-14. [Online]. Available: \url{https://www.epri.com/pages/sa/opendss}
\BIBentrySTDinterwordspacing

\bibitem{Sen}
A.~Alkhonain, A.~Singhal, A.~K. Bharati, and V.~Ajjarapu, ``Sensitivity-aware reactive power dispatch of ders to support transmission grid during emergency,'' in \emph{2023 North American Power Symposium (NAPS)}, 2023, pp. 1--6.

\bibitem{DER}
A.~Singhal, A.~K. Bharati, and V.~Ajjarapu, ``Deriving ders var-capability curve at tso-dso interface to provide grid services,'' \emph{IEEE Transactions on Power Systems}, vol.~38, no.~2, pp. 1820--1833, 2023.

\bibitem{Sen4}
Z.~Li, Q.~Guo, H.~Sun, and J.~Wang, ``Coordinated transmission and distribution ac optimal power flow,'' \emph{IEEE Transactions on Smart Grid}, vol.~9, no.~2, pp. 1228--1240, 2018.

\bibitem{opf1}
H.~Nikkhajoei and R.~Iravani, ``Steady-state model and power flow analysis of electronically-coupled distributed resource units,'' \emph{IEEE Transactions on Power Delivery}, vol.~22, no.~1, pp. 721--728, 2007.

\bibitem{opf2}
M.~Z. Kamh and R.~Iravani, ``Unbalanced model and power-flow analysis of microgrids and active distribution systems,'' \emph{IEEE Transactions on Power Delivery}, vol.~25, no.~4, pp. 2851--2858, 2010.

\bibitem{opf4}
M.~M.~A. Abdelaziz, H.~E. Farag, E.~F. El-Saadany, and Y.~A.-R.~I. Mohamed, ``A novel and generalized three-phase power flow algorithm for islanded microgrids using a newton trust region method,'' \emph{IEEE Transactions on Power Systems}, vol.~28, no.~1, pp. 190--201, 2013.

\bibitem{opf5}
A.~A. Eajal, M.~A. Abdelwahed, E.~F. El-Saadany, and K.~Ponnambalam, ``A unified approach to the power flow analysis of ac/dc hybrid microgrids,'' \emph{IEEE Transactions on Sustainable Energy}, vol.~7, no.~3, pp. 1145--1158, 2016.

\bibitem{opf7}
C.~Li, S.~K. Chaudhary, M.~Savaghebi, J.~C. Vasquez, and J.~M. Guerrero, ``Power flow analysis for low-voltage ac and dc microgrids considering droop control and virtual impedance,'' \emph{IEEE Transactions on Smart Grid}, vol.~8, no.~6, pp. 2754--2764, 2017.

\bibitem{opf8}
D.~Li, Y.~Su, F.~Wang, M.~Olama, B.~Ollis, and M.~Ferrari, ``Power flow models of grid-forming inverters in unbalanced distribution grids,'' \emph{IEEE Transactions on Power Systems}, vol.~39, no.~2, pp. 4311--4322, 2024.

\bibitem{LineDistFlow}
M.~Baran and F.~Wu, ``Optimal sizing of capacitors placed on a radial distribution system,'' \emph{IEEE Transactions on Power Delivery}, vol.~4, no.~1, pp. 735--743, 1989.

\bibitem{LineDist3Flow}
D.~B. Arnold, M.~Sankur, R.~Dobbe, K.~Brady, D.~S. Callaway, and A.~Von~Meier, ``Optimal dispatch of reactive power for voltage regulation and balancing in unbalanced distribution systems,'' in \emph{2016 IEEE Power and Energy Society General Meeting (PESGM)}, 2016, pp. 1--5.

\bibitem{Der27}
E.~Schweitzer, S.~Saha, A.~Scaglione, N.~G. Johnson, and D.~Arnold, ``Lossy distflow formulation for single and multiphase radial feeders,'' \emph{IEEE Transactions on Power Systems}, vol.~35, no.~3, pp. 1758--1768, 2020.

\bibitem{taheri2024optimized}
B.~Taheri, R.~K. Gupta, and D.~K. Molzahn, ``Optimized lindistflow for high-fidelity power flow modeling of distribution networks,'' \emph{arXiv preprint arXiv:2404.05125}, 2024.

\bibitem{interactionTSO_DSO}
S.-W. Park and S.-Y. Son, ``Interaction-based virtual power plant operation methodology for distribution system operator’s voltage management,'' \emph{Applied Energy}, vol. 271, p. 115222, 2020.

\bibitem{IEEE13}
I.~P.~E. SOCIETY, ``Ieee pes test feeder,'' \url{https://cmte.ieee.org/pes-testfeeders/resources/}.

\end{thebibliography}
\end{document}